# Chaotic Slow Slip Events in New Zealand from two coupled slip patches: a proof of concept

Thomas Poulet[1], Sandro Truttmann[2], Victor Boussange[3], Manolis Veveakis[4]

[1]CSIRO Mineral Resources, Kensington, WA, Australia

[2]Institute of Geological Sciences, University of Bern, Bern, Switzerland

[3]Swiss Federal Research Institute WSL, Birmensdorf, Switzerland

[4]Civil and Environmental Engineering, Duke University, Durham, NC, USA

**Abstract**

Recent studies showed that seemingly random Slow Slip Events (SSEs) can display chaotic patterns within the largest source of seismic hazards in New Zealand, the Hikurangi subduction zone. Some irregular SSE occurrences are therefore not arbitrary but behave with short-term predictability. However, the forecasting challenge persists as observations remain too short and noisy to constrain purely data-driven solutions, calling for a physics-based modelling approach. Here we propose a physical model of two coupled oscillators, each capturing the behaviour of a single slow-slip patch, for the deep Kaimanawa and the shallow East Coast SSEs respectively. The simplified model successfully reproduces the type of chaotic behaviour observed at the Global Navigational Satellite System station in Gisborne, yielding SSEs of appropriately varying amplitude and duration. Those results reveal that the multi-physics response of the shear zone strongly controls the underlying system, even before accounting for any geometrical complexity or distribution of material properties.

**Plain Language Summary**

The largest source of seismic hazards in New Zealand, the Hikurangi subduction zone, displays slow slip events (SSEs) lasting weeks to months and recurring every few years, releasing very large forces. Their irregularity if often mistaken for randomness, but some around Gisborne are chaotic, which means ruled by deterministic laws and therefore predictable to some extent. In practise, however, their forecast remains too difficult based on Global Navigational Satellite System observations only. Here we propose a physical model that reproduces qualitatively such chaotic SSEs. The simplified model considers the interactions between two well-known slow-slip areas, the deep Kaimanawa and the shallow East Coast SSEs, with the behaviour of each shear zone described by their linked mechanical, chemical, thermal and hydraulic responses from a material science perspective. The results show that the coupling between two tractable physicochemical responses of the shear zone controls the complex overall behaviour of the system comprising those two areas, without needing to account for the specific geometrical complexity or distributions of material properties in first considerations. This study opens the door to SSE forecasting despite data limitations, by proposing the core set of equations needed for calibrating observations, providing physical consistency to complement data-driven approaches.

**Key Points**

- Coupled oscillator models of the Kaimanawa and East Coast slip patches explain chaotic Slow Slip Events in the Hikurangi subduction zone.

- The subduction zone's chemo-physical response mainly controls its chaotic behaviour, before geometrical and spatial considerations.

- Multi-physics modelling provides a way to forecast Slow Slip Events near Gisborne, notwithstanding challenging calibration pending.

## 1 Introduction

In subduction environments, frequently occurring Slow Slip Events (SSEs) significantly contribute to the overall deformation; they release large amounts of strain (Dixon et al., 2014) and may even trigger large magnitude, destructive earthquakes (Bletery & Nocquet, 2023; Obara & Kato, 2016; Weng & Ampuero, 2022). SSEs have been identified worldwide (Weng & Ampuero, 2022) through Global Navigation Satellite System (GNSS) monitoring (Heflin et al., 2020) and their episodic behaviour seems universal. While some SSE systems reveal striking periodic recurrence intervals, e.g. in Cascadia (Rogers & Dragert, 2003), SSEs in New Zealand's Hikurangi subduction zone display much more irregular duration and recurrence times (Wallace, 2020). Interestingly, it was recently shown that SSEs can behave in a deterministic manner, not only in simpler geometrical environments like Cascadia (Gualandi et al., 2020) but also in New Zealand (Truttmann et al., 2024). Those SSEs are therefore not random but chaotic, which means that both occurrence and magnitude could be described by a mathematical system of equations and that some short-term SSE predictability is theoretically possible. However, reliable forecasts are not yet available as monitoring observations have not been operating for long enough and remain too noisy for purely data-driven approaches (Truttmann et al., 2024).

Some physical understanding of shear zones is required, in general and in subduction environments specifically, to enhance the interpretation of observations. Fortunately, such knowledge progresses fast through theoretical (Alevizos et al., 2014; Rice, 2006; Sulem & Famin, 2009), experimental (Brantut et al., 2008; Han et al., 2007; Marone, 1998), observational (Collettini et al., 2013), or numerical (Rattez et al., 2018; Veveakis et al., 2014) studies. However, no comprehensive model has yet been proposed to explain cases as complex as New Zealand and the challenge remains open to suggest an underlying system of equations controlling that system. Here, we present a first feasibility study.

## 2 Materials and Methods

### 2.1 The Oscillator Model

One multi-physics oscillator model (Alevizos et al., 2014) has proven its ability to explain single shear zone scenarios and link temporal evolution of subduction zones (Veveakis et al., 2014) with their corresponding spatial signature (Poulet, Veveakis, Herwegh, et al., 2014). The system considers a fully saturated shear zone subject to shear heating and fluid release by chemical decomposition, with a rate- and temperature-dependent frictional behaviour in a low-permeability environment (Tisato et al., 2024). The shear zone, only meters thick (Poulet, Veveakis, Herwegh, et al., 2014; Poulet, Veveakis, Regenauer-Lieb, et al., 2014) but extending over kilometres, can be modelled in one dimension (1D) across its thickness as a function of temperature ($T$) and excess pore pressure ($\Delta P$). Shear heating from the creeping fault can eventually trigger reversible endothermic chemical reactions releasing fluid, whose extent can be tracked by the corresponding partial solid ratio ($s$). The resulting system of equations also includes the effects of the resulting pore pressurisation, porosity ($\phi$) variation, and can be expressed as the evolution of two state variables, the temperature ($T$) and excess pore pressure ($\Delta P$), as functions of time ($t$) and space ($z$):

$$\begin{cases} \frac{\partial \Delta P}{\partial t} = \frac{\partial}{\partial z}\left[\frac{1}{Le}\frac{\partial \Delta P}{\partial z}\right] + (1-\phi)(1-s)\mu_r e^{\frac{Ar\delta T}{1+\delta T}} \\ \frac{\partial T}{\partial t} = \frac{\partial^2 T}{\partial z^2} + \left[Gr((1-\Delta P)_+)^n e^{\frac{\alpha Ar}{1+\delta T}} - (1-\phi)(1-s)\right]e^{\frac{Ar\delta T}{1+\delta T}} \end{cases} \quad (1),$$

where $n$ denotes the rate sensitivity of friction and the subscript "+" the positive part of a term, to account for hydrofracturing cases. The remaining dimensionless groups used include the Gruntfest number $Gr$, Lewis number $Le$, as well as the $\mu_r$, $Ar$, and $\delta$ parameters. The model is one-dimensional across the shear zone given its finite thickness of a few meters (Poulet, Veveakis, Regenauer-Lieb, et al., 2014), compared to the extension of the subduction zone in kilometres along its dip and hundreds of kilometres laterally. The values of all parameters determine the location (depth) of the oscillator.

From the resulting oscillator behaviour, we can retrieve the strain rate profile

$$\dot{\epsilon} = ((1-\Delta P)_+)^n e^{\frac{-(1-\alpha)Ar}{1+\delta T}} \quad (2),$$

which can be integrated in space across the shear zone to obtain the velocity, and then integrated in time to get the displacement.

The system of equations (1) is expressed in dimensionless form to highlight the controlling parameters, including the Gruntfest number (Gruntfest, 1963), $Gr$, defined as the ratio of mechanical work over the enthalpy of the chemical reaction, and the modified Lewis number, $Le$, inversely proportional to permeability. There exist various stability regimes (Alevizos et al., 2014) and for specific parameters the transient system exhibits a limit cycle leading to slip events lasting a few days – slow slip corresponding to observed seismic tremors – and occurring periodically every few months (Figure 1).

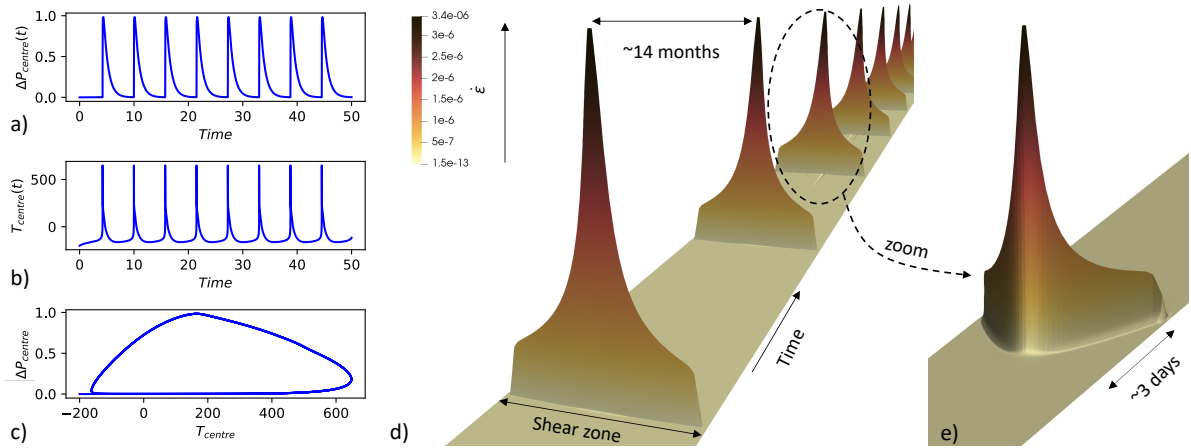

*Figure 1: Response of a single multi-physics serpentinite oscillator modelling a 1D chemical shear zone through a few stick-slip cycles. The model is based on two primary dimensionless variables, the temperature (T) and excess pore pressure (ΔP), following the system of equations 1, with parameter values from Table SI1. Time evolution at the centre of*

*the shear zone of (a) the excess pore pressure $\Delta P_{centre}$ and (b) the temperature $T_{centre}$. (c) Plotting $\Delta P_{centre}$ vs $T_{centre}$ shows the periodic limit cycle of the oscillator. (d) Profile evolution of strain rate $\dot{\epsilon}$, using Equation 2, showing strong localisation at the shear zone centre. Slip events occur every 14 months and last for a few days.*

Such a single oscillator is applicable to scenarios with enough geometrical simplicity for the 1D assumption to hold, like the Cascadia subduction zone (Alevizos et al., 2014), and can explain observed periodically recurring SSEs, called Episodic Tremor and Slip (ETS) events, occurring in a specific depth range (Rogers & Dragert, 2003). The multi-physics (thermo-hydro-mechanical-chemical) nature of the system encompasses various behaviours including thermal dehydration (Sulem & Famin, 2009), weakening (Rice, 2006), pressurisation of pore fluid (Passelègue et al., 2020), and healing (Shreedharan et al., 2023). The resulting oscillatory regime is confined to specific parameter ranges of the dimensionless parameters (Alevizos et al., 2016). While previous analyses have mostly focused on the low permeability threshold below which oscillations occur (Poulet, Veveakis, Regenauer-Lieb, et al., 2014), numerical analyses for specific parameter sets can establish quantitative bounds (e.g., Figure SI1). The definition of these parameters offers a direct connection to physical properties, thereby enabling the prediction of specific depth ranges where SSEs can occur.

## 2.2 New Zealand: Two Coupled Oscillators

In New Zealand, SSE patterns are much more complex and not as periodic as in Cascadia (Wallace, 2020), likely related to the more heterogeneous structure of the Hikurangi subduction zone. Yet, these SSEs also exhibit characteristics of a chaotic system, raising hope for short-term predictability (Truttmann et al., 2024). However, forecasts with purely data-driven methods require clean signals and long measurement periods, which are not available yet; this is where physics can help.

The curved geometry of the Hikurangi Trough off New Zealand marks the Southern end of the Kermadec and Tonga plates, which extend rather linearly over more than 2,000km northward (Figure 2b). Hence, we assume a 1D shear zone configuration away from the Southern end and focus on the northern-most SSEs in New Zealand, the Deep Kaimanawa SSEs and East Coast SSEs (Figure 2a) but ignore the Kapiti SSEs further South, since the 1D-assumption does not hold in the southern part of the Hikurangi subduction zone anymore due to its bending, adding further complexity (Lee et al., 2024). SSEs might occur along the whole Kermadec and Tonga plates but cannot be detected since no offshore

GNSS data is available. Based on cumulative slip observations (Wallace, 2020), we consider a conceptual cross-section around Gisborne (Figure 2c) and interpret the presence of SSEs in two localised patches at different depth levels as two separate oscillator cells. For this conceptual analysis, we omit the question of rock composition in each patch and use a previously published oscillator based on carbonates for both oscillators (Poulet, Veveakis, Herwegh, et al., 2014) (Table SI1), noting the flexibility of the dimensionless approach to calibrate the system for other conditions.

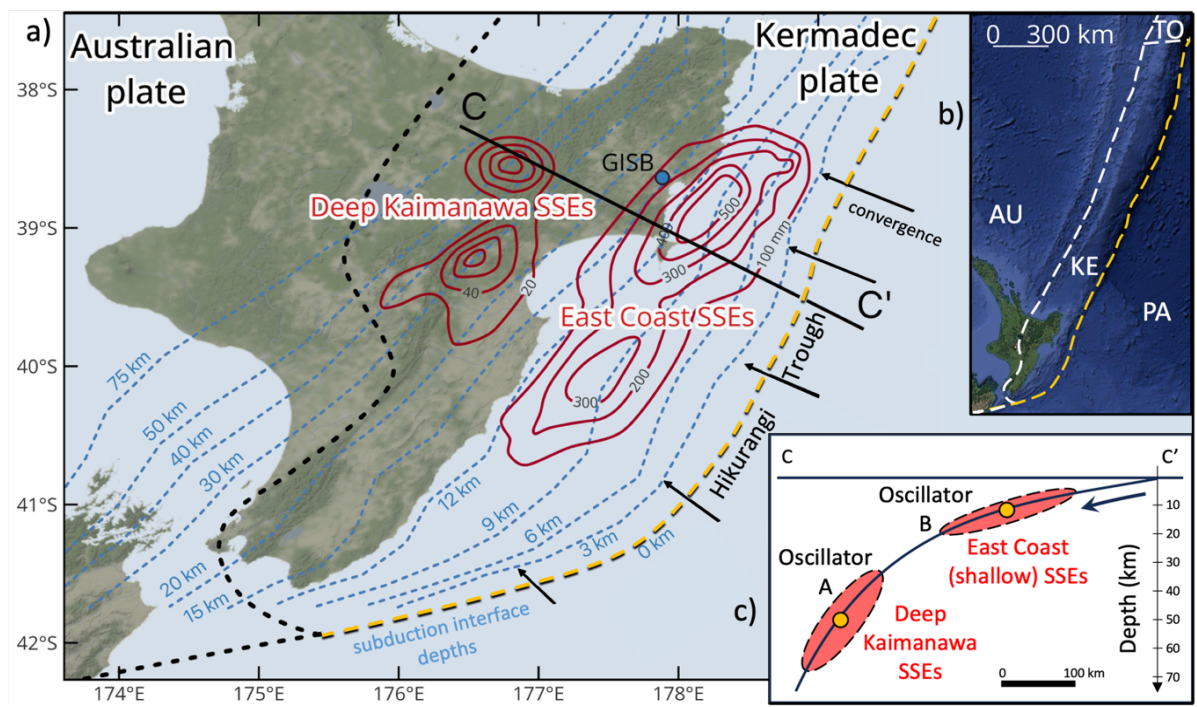

*Figure 2: Selected Slow Slip Events in New Zealand's North Island. (a) Map showing cumulative slip contours (2002-2014) from the Deep Kaimanawa and East Coast SSEs as well as plate boundaries (black dashes), Hikurangi Trough (yellow dashes), subduction interface depths (Wallace, 2020), and location of the Gisborne station (GISB, blue marker). (b) Larger scale view of the Kermadec (KE), Australian (AU), Pacific (PA) and Tonga (TO) plates, with plate boundaries (white dashes) and Hikurangi Trough (yellow dashes); The elongated profile of the Kermadec plate justifies ignoring geometry, at first order, on any subduction profile not too close to the plate's lower edge. c) Schematic vertical cross-section across the deep and shallow SSEs. Each slip patch is modelled by an oscillator.*

The two oscillators correspond to slip patches at different depths for the same subduction zone and are geomechanically linked. Their exact interaction remains an open problem but interactions between the deep and shallow part of a slab have been observed in subduction environments (Gardonio et al., 2024), with synchronisation between shallow and deep seismic activities (Bouchon et al., 2022; Jara et al., 2017) pointing to the role of rigidity (Sallarès & Ranero, 2019). Here, we aim to demonstrate the potential of a physical approach to capture complex SSEs patterns and model this SSE system by coupling the two oscillators, postulating that it is that coupling which controls its chaotic response (Veveakis et al., 2017). Any slip event from one oscillator affects the other, mainly through the force chain along the connecting subduction zone. As the driving system arguably stems from greater depth and effects propagate upwards (Poulet, Veveakis, Regenauer-Lieb, et al., 2014), we herein consider a one-way coupling only, where the slip events of the deep oscillator trigger excess forces on the shallower one, which can be simulated in a conceptual manner through the evolution of its Gruntfest number. This idealised coupling aims at identifying the responsible driving factor while avoiding the complex physical and geometrical considerations of the real configuration.

In that respect, we expect the two oscillators to communicate through the multiphysical response of the connecting domain. This domain is, in principle, experiencing mechanical (velocity and displacement), thermal and pore pressure gradients. In the present approach, all these fields are coupled (see Eq. 2), thereby suggesting that linking the two oscillators could be performed directly through their boundary conditions, and in particular through the temperature field since excess pore pressure is equal to zero everywhere but inside the shear zones. As such, the simplest approach consists of coupling the two oscillators with a thermoelastic spring, assuming the two slip patches are connected in series and share the same stress $\sigma = E(\varepsilon + \alpha \Delta T)$ through a thermoelastic connecting domain, where $\alpha$ denotes a thermal expansion coefficient and $E$ an elastic modulus. The stress $\sigma$ can then be calculated as $\sigma = (E_1 + E_2)\varepsilon + (E_1 \alpha_1 \Delta T_1 + E_2 \alpha_2 \Delta T_2)$. This, in turn, means that stress perturbations under fixed displacement (constant strain) are driven by thermal stresses alone, which is a constrain of the current approach and should be generalized in the future. In the example of two Coulomb-type faults with friction coefficient μ, connected in series through a thermoelastic half space under fixed displacement, the applied shear stress change would then be $\Delta \tau = \tau - \tau_0 = \mu \Delta \sigma = \mu(E_1 \alpha_1 \Delta T_1 + E_2 \alpha_2 \Delta T_2)$.

Based on these considerations, we take a proxy for the force transfer between connected oscillators through the evolution of the Gruntfest number (Veveakis et al., 2017), $Gr$. Since the strain rate is strongly temperature-controlled, we update $Gr$ for one oscillator linearly with the temperature derivative of the other, as a proxy for slip events. Setting an activation value $\delta T_{min}$ and a maximum bound $\delta T_{max}$ for $\frac{\partial T_1}{\partial t}\big|_{centre}$, we compute the proportionality coefficient as

$$\beta_1 = \left(\frac{\partial T_1}{\partial t}\bigg|_{centre} - \delta T_{min}\right)/(\delta T_{max} - \delta T_{min}) \qquad (3),$$

where $\delta T_{min}$ and $\delta T_{max}$ can be precomputed since the first (external) oscillator is unperturbed (one-way coupling). The $Gr$ parameter for the second (internal) oscillator can be expressed as

$$Gr_2 = Gr_{2_0} + \beta_1 \Delta Gr \qquad (4),$$

where $\Delta Gr$ is taken as a constant parameter, whose value is investigated in the study based on the resulting patterns obtained.

This approach allows to retrieve a variety of responses (Heltberg et al., 2021), including quasiperiodic dynamics when oscillators do not lock, entrainment with new period for small coupling strengths, period-doubling dynamics, and chaotic dynamics (Figure 3a). The high sensitivity of the parameters (e.g., Figure 3b) places the full picture of the various regimes beyond the scope of this study. We consider the displacement of the shallow oscillator as proxy observation to compare against GNSS data from the Gisborne station (GNS Science, 2000), which records the shallow East Coast SSEs exceptionally well (Figure 4a).

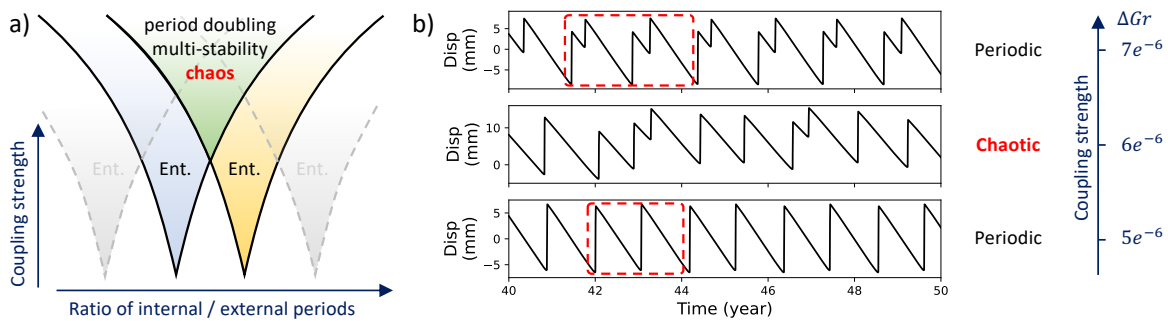

*Figure 3: Possible responses of one-way coupling for two oscillators. (a) Schematic diagram illustrating the variety of responses possible (Heltberg et al., 2021), including quasi-periodicity when oscillators do not lock (outside coloured areas), entrainment regions ("Ent."), known as Arnold tongues (Arnol'd & Avez, 1968), with new periods for small coupling strength, multi-stability, period-doubling, or chaotic dynamics for specific values of parameters. (b) Real example responses for our coupled oscillators, showing the sensitivity of the coupling strength and small parameter ranges leading to chaotic responses of fault slip at the shallow oscillator, with $\Delta Gr$ defined in Equation 4. Red dashed windows indicate periodic patterns.*

## 3    Results - Chaotic Displacement Patterns

The results show that we can recover a chaotic response of the system (Figure 3b), leading qualitatively to the displacement patterns observed, with instant slip events of various amplitudes and durations. By definition of a chaotic system, its response is fairly sensitive to input parameters and more periodic regimes can also be retrieved (Figure 3) as expected (Heltberg et al., 2021). With the proof of concept established, we further investigate the parameter space. From a manual analysis, we show that other parameters allow to reproduce displacement results that are qualitatively close to real observations, with series of small temperature jumps interrupted by larger and longer jumps (Figure 4b). The comparison of the simulation results with observations remains conceptual, since the prototype nature of the one-way coupling used would not warrant the considerable time required to calibrate such a complex and sensitive model. As such, only the temperature evolution is displayed (Figure 4b), to emphasize its role as the driving parameter responsible for the slip patterns, which would require extra parameter tuning to directly match observations.

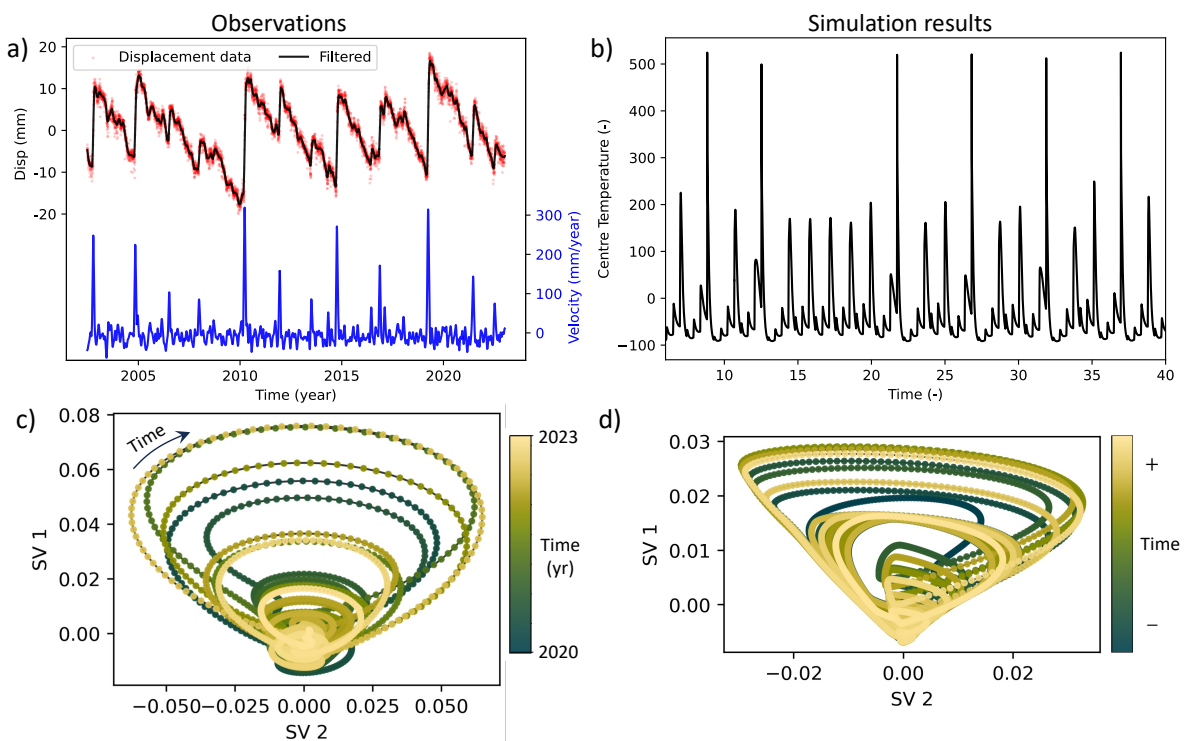

*Figure 4: Qualitative comparison of observations from the GISB station (left) and the results of two coupled oscillators (right). (a) GNSS data from the GISB station, including daily positions (red dots), smoothed displacement (black) using low-pass Finite Impulse Response (FIR) filtering (Truttmann et al., 2024, see Appendix) and corresponding slip velocity (blue). (b) Chaotic model results of dimensionless temperature evolution at the centre of the shear zone. (c) Phase space representation of the first two singular vectors (SV) derived by singular system analysis of GISB observations. Trajectories*

*follow large cycles during SSEs and remain close to the origin in between (inter-SSE), whether as signal noise or small amplitude slip events. (d) Phase space representation of chaotic model results, displaying similar characteristics as for GISB observations. Here, the small inter-SSE cycles are not noise, given the noise-free simulation results.*

## 4  Discussion

It is particularly exciting to use a physical system and retrieve signals that bear the characteristics of GNSS observations. We therefore argue that the oscillator model can not only explain the complex mechanisms at play, but also provide a potential tool to forecast irregular SSEs in New Zealand. Indeed, if the physical parameters can be calibrated to match observations, then the phase space analysis (Truttmann et al., 2024) can be performed on the corresponding clean signal from the physical system that could lead to the maximum theoretical predictions.

Despite the missing calibration step, we rebuild a multi-dimensional phase space of the results(Broomhead & King, 1986), which describes all possible states of the dynamical system, as was done from the GNSS data (Truttmann et al., 2024). The resulting structure observed in phase space (Figure 4d) matches qualitatively that from GNSS observations (Figure 4c) and reveals the underlying attractor for those trajectories (Abarbanel, 1996). The critical difference is that we now have a description of the physical system involved. In particular, the presence of numerous cycles of smallest amplitude – corresponding to the inter-SSE signal (Truttmann et al., 2024) – indicate that there might be a stronger signal component than previously thought in those parts of the data usually thought of as noise, as previously proposed (Frank, 2016; Jolivet & Frank, 2020; Kato & Nakagawa, 2020; Truttmann et al., 2024). This also increases optimism for purely data-driven analyses.

In our case, the smallest jumps (Figure 4b) remain quasi-periodic because of the one-way coupling. Their occurrence could be perturbed by introducing the reciprocal coupling of the shallow oscillator impacting the deep one. This added complexity should be considered for realistic calibration attempts but goes beyond the scope of this study.

This study suggests the feasibility of using coupled slip oscillators to model New Zealand SSE characteristics, with potential applications to other SSEs globally, if several slip patches have been identified and a chaotic signal suspected. Amongst a growing trend of data-driven methods being deployed to analyse complex SSE observations (Corbi et al., 2020; Gualandi et al., 2020, 2024; Johnson & Johnson, 2024; Truttmann et al., 2024), the complementarity of physics-based approaches remains key to identify controlling factors and explain the processes involved. The proposed oscillator model (Veveakis et al., 2014) can describe episodic SSEs, their timing (Figure 1d), duration (Figure 1e), spatial signature (Poulet, Veveakis, Herwegh, et al., 2014) and the underlying reason for the number and location of slip patches observed (Figure 2). Interestingly, it can also apply to one-off earthquakes (Veveakis et al., 2014) in similar environments, for which refined data processing reveals precursors with the same exponential build-up (Bletery & Nocquet, 2023). The fact that an approximate one-way coupling of two oscillators can explain the chaotic behaviour and qualitatively match observations suggests that the physics of the shear zone really is the main control of the overall system behaviour, despite all the complexity not currently accounted for and which clearly plays a role, including 3D geometry and rheological heterogeneities of the subduction interface (Perez-Silva et al., 2022; Weng & Ampuero, 2022). This analysis helps balance the potentially excessive importance granted to geometrical factors (Lee et al., 2024).

This study highlights the effectiveness of simplified modelling, which should not suffer from our ability to simulate more realistic 3D problems. The homogenisation of an entire slip patch with a single oscillator aims at capturing the essence of the overall system instead of every aspect. The suggested prototype can and should eventually be extended with extra features for better applicability, including the continuous behaviour along the shear zone, 3D geometrical effects, two-way couplings, and more accurate non-instantaneous force coupling, including with other slip cells. Yet, the physical control of fluid-release reactions on complex SSEs at the large scale raises confidence that simplified underlying equations can be derived to simulate what is otherwise one of the most complex geological scenarios. Also, the analysis of the generated displacement signal in phase space (Figure 4d) highlights that small amplitude parts of observed signals are not necessarily noise or errors, which warns against too much smoothing on real-world SSE signals.

## 5    Conclusions

The characterisation of a complex subduction zone by the simple coupling of multi-physics oscillators demonstrates that complex SSE dynamics could be described by a physical system, implying exciting prospects in terms of SSE predictability. While nearly periodic SSE signals can intuitively be accepted as deterministic, like in Cascadia (Gualandi et al., 2020), it is fascinating to unravel the physical processes behind irregular signals from

New Zealand equally identified as deterministic systems with chaotic dynamics (Truttmann et al., 2024). Modelling the coupled force transfer interactions between two slip patches could explain chaotic SSEs in the Hikurangi subduction zone and potentially other settings, since various types of fault slip observations seem to arise from the same physical mechanisms (Passelègue et al., 2020). Given the remarkable qualitative match between empirical data and the derived physical model, the latter could be calibrated with data-inference methods (Boussange et al., 2022) to provide short-term predictions. Furthermore, the parameterization of dimensionless groups with a neural network could correct unresolved processes (Beucler et al., 2024; Rasp et al., 2018), especially for the coupling, and additionally be used to improve our fundamental understanding of physical processes by an offline analysis of the learnt parametrization (Rackauckas et al., 2020). Alternatively, the physical processes that we have identified could be used to constrain data-driven approaches following a physics-informed neural network approach (Lagergren et al., 2020; Yazdani et al., 2020), mitigating the problem of data deficiency. In either case, by harnessing physical knowledge of the underlying systems, these approaches could leverage the available data and greatly improve our capacity to forecast SSE and their fast, potentially destructive counterparts (Dixon et al., 2014; Obara & Kato, 2016; Uchida et al., 2016).

## 6  Appendix – Phase Space Reconstruction

Time series can conveniently be analysed using Singular Spectrum Analysis (Broomhead & King, 1986), which decomposes the trajectory matrix X, reconstructed from M lagged copies of the signal, into a sum of Singular Vectors (SV), similar to an eigenvalue decomposition. The SVs then represent the new coordinates in phase space, which is a frequently used nonlinear analysis tool to study system dynamics. In this study, we follow the approach detailed in a previous publication (Truttmann et al., 2024) and show 2D representations of resulting phase spaces (Figure 4), where the values of the first two vectors are used as coordinates, varying with time. For real observations (Figure 4c), raw data is herein smoothed by applying a low-pass finite impulse response (FIR) filter utilising a Hamming window, with a cut-off frequency of 1/50 and a window span of 60 days (Truttmann et al., 2024). Smooth data obtained from simulations can be transformed into phase space directly (Figure 4d). We then reconstruct the phase space coordinates using a window length of $M = 60$.

## 7 Code Availability

The code used to model the coupled oscillators is available (Boussange et al., 2024) at
https://zenodo.org/doi/10.5281/zenodo.13300509

## 8 Data Availability

The GNSS data used in this study is available from GNS Science (GNS Science, 2000) at
https://doi.org/10.21420/30F4-1A55